\documentstyle[12pt]{article}

\begin{document}

\title{\normalsize RELATIVISTIC DEUTERON STRUCTURE FUNCTION AT LARGE $Q^2$}

\author{ \underbar{\bf J. Paulo Pinto}, {\bf A. Amorim} and {\bf F. D. Santos}
\\
{\centerline {\footnotesize  {Centro de F\'{\i}sica Nuclear da Universidade de Lisboa
 Av. Gama Pinto, 2, 1699 Lisboa , Portugal.}}}
}
\maketitle

{\bf E--mail}: jpinto@alf1.cii.fc.ul.pt

{\bf fax}: 351 1 7954288

\begin{abstract}
The deuteron deep inelastic unpolarized structure function $F_2^D$ is calculated
using the Wilson operator product expansion method. The long distance behaviour,
related to the deuteron bound state properties, is evaluated using the 
Bethe--Salpeter equation with one particle on--mass--shell. The calculation of
the ratio $F_D^2/2F_N^2$ is compared with other convolution models showing important
deviations in the region of large $x$. The implications in the evaluation of the 
neutron structure function from combined data on deuterons and protons are discussed.
\end{abstract}

\vspace{3cm}


{\bf keywords}: \,Deuteron; neutron; Structure functions; Binding; Relativistic effects

\newpage

\section{Introduction}
\label{sec1}

Since the discovery of the EMC~\cite{a1} effect it became clear that the nucleon structure
functions inside nuclei are different from the ones determined for free nucleons. 
In the quark--parton model point of view, this means that
the quark and the anti--quark sea
distributions, in the nucleon, are changed by the surrounding nucleons.
Traditional methods~\cite{a2,a3} that considered only Fermi motion of the nucleons  
are not sufficient to describe the experimental data. Several efforts have been
done to explain this deviation, mainly along two different approaches.
The first is QCD motivated~\cite{a4} and assumes a $Q^2$ rescaling, i.e.,
$F_2^A(x,Q^2)/A=F_2^N(x,\xi Q^2)$, where $\xi $ is a A--dependent
free parameter determined by the best fit to the data. The other typical approach \cite{a5}$- 
\!\!$\cite{a11a} is based on the nuclear degrees of freedom namely, 
the nucleon--nucleon interaction
involving mesonic exchange contributions and other nuclear structure effects.
In this work special attention is given to the method described 
in Ref.~\cite{a10,a11}, which relies on the Wilson operator
product expansion (OPE). This is a systematic and 
self--consistent approach inspired in nuclear physics concepts which will be used
here in the calculation of the deuteron structure function $F_2^D$.

 A comparison~\cite{a12} between the  $Q^2$ rescaling model and the OPE model
 through the moments $M_n^A(F_2^A)$ of the structure functions,
 shows that both methods agree in the intermediate region of $x$ $(0.2<x<0.7)$
 corresponding to small values of $n$. For
large $n$, it seems that the OPE is more appropriate since
it describes the experimental data in the region of large $x$. The 
phenomenological nature of the $Q^2$ rescaling method results from the fact
that it is hard to predict the value of $\xi $ within QCD.
On the other hand, the OPE method has no free parameters and, in particular, the deuteron
structure function can be obtained in a consistent way.

Ref.\cite{a11a} provides a general covariant method to describe deep--inelastic
scattering (DIS) that does not rely  on the Bjorken
limit. In this model the deuteron structure function depends upon the
off--shell nucleon structure function and the Bethe--Salpeter deuteron wave function.
However, there is no factorization, which implies that, in general, the convolution formula is
not valid. The usual ambiguities related to the off--shellness problems in
DIS are not present, but the nucleon distribution
function, which is an important tool for the interpretation of
other processes involving the deuteron,
is not consistently obtained within this formalism. 

Some important considerations about parton distribution functions and QCD analysis of
DIS  rely on the experimental structure functions of the nucleons.
 In general, the neutron structure
function is extracted from combined experimental data on the proton and deuteron~\cite{a13},
taking into account only the Fermi motion of the nucleons in the deuteron. The question
of whether the EMC effect is a feature only of the heavy nuclei or should also be
considered as leading to an important correction in light nuclei
must be discussed. In principle one should
take into account the binding effects and the meson exchange currents, required to describe
the EMC effect in heavy nuclei, also in light nuclei. It has been shown ~\cite{a14}
that the additional corrections are important in the deuteron and that
the extracted neutron structure function is model dependent due to the nuclear contributions.
This is particularly important at large $x$ where the discrepancy with the
 data is of the order $\simeq 25$\% and exceeds the
experimental error bars~\cite{a13}. It is thus clear that, from a nuclear physics point of view,
one should consider corrections to the Fermi motion of the nucleons even
for light nuclei.

In this paper the deuteron structure function is calculated using a relativistic
formalism, based on the Wilson operator product expansion method,
following Ref.~\cite{a10,a11}. The deuteron
 Bethe--Salpeter amplitudes are obtained from
a quasi--potential equation with one particle on--shell ~\cite{a15}. 
Section \ref{sec2} is an overview of the formalism and presents the
relevant expressions used in the calculation.   The
main features of the relativistic deuteron wave function used in our calculations
are described in section \ref{sec4} and in section \ref{sec3} we discuss
the results of the ratio $F_2^D/2F_2^N$ through a comparison with other approaches.

\section{The Formalism}
\label{sec2}

The basic feature of the Wilson operator product expansion is the possibility of
factorizing the DIS amplitude into two pieces in the limit  $Q^2 \to \infty$,
 one associated with  the long distance
behaviour and the other with the short distance behaviour.
The pertubative QCD formalism determines the large $Q^2$ behaviour of the amplitude.
On the other hand, a nuclear physics approach can be used to deal with the target nuclei, where
the long distance behaviour can in general be predicted from the bound state properties.

According to the OPE method~\cite{a10,a11} the deuteron structure 
function $F_2^D$ satisfies the relation
\begin{equation}
\int_{0}^1 dx\,x^{n-2} F_2^D(x,Q^2)=\sum_{a} C_{a,n-1}^{(2)}(Q^2)\, \mu_n^{a/D}\,,
\label{eq1} \end{equation}
where $x=Q^2/2P_D.q$ is the usual Bjorken scaling variable. 
The sum in the second member of Eq.(\ref{eq1}) runs over different 
fields of the theory, according to the twist--2
approximation. The Wilson coefficients $C_{a,n}^{(2)}$ are target independent,
giving the short distance behaviour of $F_2^D$. In the impulse approximation, it is found that  
these coefficients are identical to the moments of the structure function $F_2^a$ of
nucleons $(a=N)$ or mesons $(a=B)$,
\begin{equation}
 C_{N,n}^{(2)}(Q^2)=M_n(F_2^N) \hspace{2cm} C_{B,n}^{(2)}(Q^2)=M_n(F_2^B)\,. \label{eq1a}
\end{equation}
Noting that the moments are defined as 
\begin{equation}
M_n(F)=\int_0^1 dx\,x^{n-1} F(x,Q^2)\,, \label{eq1b}
\end{equation} 
we obtain from Eqs.(\ref{eq1}),(\ref{eq1a})
\begin{equation}
M_n(F_2^D)=M_n(F_2^N)\,\mu_{n+1}^{N/D}+M_n(F_2^B)\,\mu_{n+1}^{B/D} \,,\label{eq2}
\end{equation}
where $\mu_{n+1}^{a/D}$ is interpreted as the moment of an effective distribution function
of nucleons or mesons in the deuteron. In the leading twist approximation, these functions
are give by
\begin{equation}
<\,P_D\,|\,O_a^{\mu_1 \dots \mu_n}\,|\,P_D\,>=P_D^{\mu_1} \dots P_D^{\mu_n}\,\mu_n^{a/D}
 \label{eq2a}
\end{equation}
where $O_a^{\mu_1 \dots \mu_n}$ is the set of local operators that provide the basis
for the operator product expansion.
The matrix elements from Eq.(\ref{eq2a}) can be explicitly evaluated using
the Mandelstam method ~\cite{a16} and in particular, the 
following result holds in the impulse approximation
\begin{eqnarray}
\mu_n^{N/D}&=&\frac{1}{2 (P_D.\epsilon)^n} \int \frac{d^4k}{(2\pi )^3} 
\tilde{\chi}(p_1,p_2) S(p_1) \rlap/\epsilon S(p_1) \chi(p_1,p_2) 
 S^{-1}(p_2)\,\nonumber \\
&\times & \delta\left(p_2^2-m_N^2\right)\, (p_1.\epsilon)^{n-1}\,+
\,(1 \leftrightarrow 2)  \label{eq3}
 \end{eqnarray}
where $p_1$, $p_2$ are the nucleon momenta in the deuteron and $k=(p_1-p_2)/2$. The vector
$\epsilon $ is chosen in such a way that $\epsilon^2=0$ and $\vec{\epsilon}\propto \vec{q}$ where
$\vec{q}$ is the photon momentum with
$q^2=-Q^2$. In particular, if the transfered 3--momentum $\vec{q}$ is along the z--axis one
can choose $\epsilon=(1,0,0,-1)$. This prescription is used in extracting
$\mu_n^{N/D}$ from the twist--2 operator expansion.
 
Expression (\ref{eq3}) is similar to the one presented in Ref.~\cite{a11}, but
includes a  $\delta$--function in the integrand which means that the distribution
function moment $\mu_n^{N/D}$ is calculated with one particle on--mass--shell. The difference is
related to the fact that the Bethe--Salpeter amplitudes $\tilde{\chi}$, $\chi $ 
used in this work, are determined by a quasi--potential equation with one particle
on--mass--shell~\cite{a15}. This approximation 
includes some off-shell effects (the other particle is completely 
off--shell) and it reproduces successfully the bound state properties including
the deuteron form factors~\cite{a17}.

Applying the inverse Mellin transform to Eq.(\ref{eq2}), defined as
\begin{equation}
F(x)=\frac{1}{2\pi i} \int_{\gamma-i\infty}^{\gamma+i\infty}dn\,x^{-n}M_n(F)\,, \end{equation}
 and taking into account the explicit
expression for $\mu_n^{N/D}$ one recovers the convolution formula
\begin{equation}
F_2^D(x)=\int_x^1 dy\,f^{N/D}(y)\,F_2^N(x/y)\:+\:\mbox{MEC} \label{eq4}
\end{equation}
where

\begin{eqnarray}
f^{N/D}(y)&=&\frac{i}{2 P_D.\epsilon} \int \frac{d^4k}{(2\pi )^3} \tilde{\chi}(p_1,p_2) S(p_1)
\rlap/\epsilon S(p_1) \chi(p_1,p_2) 
S^{-1}(p_2)\, \nonumber \\
&\times &\delta\left(p_2^2-m_N^2\right)\,\left\{\theta(p_1.\epsilon)
\,\delta\left(y-\frac{p_1.\epsilon}{P_D.\epsilon}\right)
\theta(-p_1.\epsilon)
\,\delta\left(y+\frac{p_1.\epsilon}{P_D.\epsilon}\right)\right\}\nonumber \\
&+&(1 \leftrightarrow 2)
 \label{eq5} \end{eqnarray}

\noindent is the distribution function of nucleons in the deuteron. The meson exchange
currents (MEC) also contributes to the deuteron structure function and the corresponding
distribution function  can be found in an analogous way~\cite{a11}. 
 
Notice that $f^{N/D}$ given by Eq.(\ref{eq5})
takes into account the binding and off--shell effects in the deuteron, including the features
needed to describe the EMC effect. It can be shown, 
using the Bethe--Salpeter normalization
condition, that the distribution function satisfies the baryonic number relation,
\begin{equation}
\int_0^1 dy\,f^{N/D}(y)=2 \,.\label{eq6}
\end{equation}
The average value of the momentum carried by the nucleons is,
\begin{equation}
<y>= \int_0^1 dy\,y\,f^{N/D}(y)\,, \label{eq7}
\end{equation}
which can be written in the form
\begin{equation}
<y>=1-\delta_N \,,\label{eq8}
\end{equation}
where $\delta_N$ is that part of the deuteron momentum carried by the mesons. The value
of $\delta_N$ controls the magnitude of the  binding effects. Since the function
 $f^{N/D}(y)$ is strongly peaked at $y=1/2$, in a first approximation one has
$f^{N/D}(y)\simeq 2 \delta(y-1/2)$, resulting in $\delta_N=0$. In this case
 the deuteron is assumed to be made of two non--interacting quasi--free nucleons,
where binding effects are neglected. This approximation leads to $F_2^D(x)\simeq 2\,
F_2^N(2x)$, and corresponds to a non--smearing motion of the nucleons. A better 
estimative of $F_2^D$ can be obtained if we expand $F_2^N$ around $y=<y>/2$ in the integrand
of Eq.(\ref{eq4}). The result is given by
\begin{equation}
F_2^D(x)=2 F_2^N(2x/<y>)+\frac{1}{2}(<y^2>-<y>^2/2) \frac{\partial^2 F_2^N(x/y)}{\partial^2 y}
|_{y=<y>/2}+ \dots  \label{eq9} 
\end{equation}

\section{Deuteron vertex}
\label{sec4}

In relativistic field theory we can describe the deuteron by the 
Bethe--Salpeter (BS) vertex $\chi_{s}$\,,where $s=-1,0,1$ is the deuteron helicity.  
This function is the fourier transform
of matrix elements involving the nucleon fermionic fields,
\begin{equation}
\chi_{s}=\int d^4x\, e^{ik.x}
<0\mid\:T\Psi(x/2)\, \Psi(-x/2)\mid\:P_D,s>  \label{eq:1}
\end{equation}
and satisfies the homogeneous BS equation \cite{r14}. It depends
on the deuteron total momentum $P_D$ and on the nucleons relative momentum $k$.
According to the Gross on--mass--shell prescription \cite{a15} this
vertex can be parameterized by,
\begin{eqnarray}
 \chi_s &=&F\rlap/\xi+\frac{G}{m_N}(k.\xi)+ \nonumber \\
&+&\frac{(\rlap/P_D/2-\rlap/k-
m_N)}{m_N}(H\rlap/\xi+\frac{I}{m_N}(k.\xi))\,, \label{eq:4}
\end{eqnarray}
where $\xi_{\mu}(P_D,s)$ are the deuteron polarization functions,
 a set of spin dependent space--like
orthogonal vectors satisfying the following identities
\begin{eqnarray}
& & \xi_{\mu}(P_D,s)\,P_D^{\mu}=0 \hspace{1.2cm} \xi^{\ast}_{\mu}(P_D,s^{'})\,
\xi^{\mu}(P_D,s)=-\delta_{s^{'} s}   \nonumber \\
& & \label{eq:3} \\
& & \sum_{s} {\xi^{\mu}}^{\ast}(P_D,s)\,\xi^{\nu}(P_D,s)=-(\,g^{\mu \nu}-
\frac{P_D^{\mu}\,P_D^{\nu}}{m_{D}^{2}}\,).  
\nonumber
\end{eqnarray} 
The scalar amplitudes $F$, $G$, $H$ and $Y$ depend only on the variable
$a=\left[(P_D.k)/m_D\right]^2-k^2$ and are determined covariantly by solving
the BS equation with one particle on--mass--shell. The solutions can be found in Ref.\cite{r15}.

The NN interaction used involves the meson exchange $(\sigma ,\omega ,\pi , \rho)$
with the parameterization taken from the first article in
Ref.\cite{a15}. We also consider two different
couplings for the pion namely, a pseudovector (PV) and a mixed 
pseudoscalar--pseudovector (PS--PV) coupling, which is characterized by the
parameter $\lambda \in [0,1]$ defined in the pion--nucleon vertex as 
\begin{equation}
g_{\pi}\left(\lambda \gamma^{5}+(1-\lambda)\frac{\rlap/q}{2 m_N}\gamma^{5}\right)
\label{eq:6}
\end{equation}
where $g_{\pi}$ is the pion coupling constant. The PV coupling leads to a
deuteron wave function consisting only of S and D states. The percentage of
P states is negligible for small $\lambda $ but increases with 
$\lambda $ and for the PS--PV coupling corresponding to $\lambda=0.4 $ 
is $0.45\,\%$. In this case the interference of the P and S states can lead to
a significant effect. In our calculations we  consider two solutions, one obtained
with $\lambda=0$ (BS0) and the other with $\lambda=0.4$ (BS04).
Both solutions are consistent with the deuteron static properties.

\section{Results and Conclusions}
\label{sec3}

The extraction of the neutron structure function from combined experimental data
on the proton and deuteron 
can now be discussed. The formalism presented here indicates that
one should use Eq.(\ref{eq4})--(\ref{eq5})
to analyse this problem consistently in the large $Q^2$ 
region where the Wilson operator product expansion is reliable. 
 The on--shell condition given in Eq.(\ref{eq3}) corresponds to a  well established and
successfull procedure in nuclear physics.
In order to get some physical insight into the corrections introduced by Eq.(\ref{eq5})
one has to compare the results obtained
by different approaches. In Fig.(\ref{fig1}) we present the results of
the ratio $F_2^D/2F_2^N$ as a function of $x_N=\left(m_D/m_N\right) x$, 
for the Bethe--Salpeter formalism with a simpler deuteron model (BSsm)
\cite{a11}, a calculation using a non--relativistic deuteron wave function
 (NR) \cite{a10}, a light--cone calculation (LC) \cite{a19}
and finally the Bethe--Salpeter (BS)
formalism with one particle on--mass--shell. In the latter calculations we used
the convolution formula (\ref{eq4}) with the corresponding distribution function
given by Eq.(\ref{eq5}), while in the other approaches the results were obtained
using the expansion (\ref{eq9}). 
The nucleon structure function $F_2^N$ is taken from Ref.\cite{a12}. 

It is clear that for small values of $x_N$ all different approaches agree. However, for
$x_N>0.3$ the deviation between the different models increases. The main reason for these 
differences can be attributed to binding effects. In fact, the value of $\delta_N$ 
differs substantively in all approaches. It goes from $\delta_N=0$ in LC calculations
to $\delta_N\simeq 5.0\times 10^{-3}$ in NR calculations with the intermediate value
$\delta_N\simeq 3.9\times 10^{-3}$ for the BSsm. In our calculations we obtain 
$\delta_N\simeq 7.2\times 10^{-3}$ and $\delta_N\simeq 6.3\times 10^{-3}$ for the
BS04 and BS0 models, respectively. 

The deviation between the two curves BS0 and BS04 
is an effect of the deuteron P states, which are present only 
on the last one. In particular, the BS04 result deviates from all the 
other calculations even for small values of $x$. This is not surprising since none of those
models consider the presence of deuteron P states. The present calculations indicate 
that there is an ambiguity in the extraction of the neutron 
structure function from the combined deuteron and proton data.
The procedure is model dependent and particularly sensitive to the 
features of the deuteron wave function. 

\section*{Acknowledgement}
We are thankful to Dr. L. P.  Kaptari and Dr. C. Louren\c co
for useful discussions.
This work was partially supported by JNICT, Grant FMRH/BD/909/93.

\bibliographystyle{revc}

\begin{figure}[h]
\caption{Theoretical predictions for the ratio $F_2^D/2F_2^N$ using four 
different approaches. The curves
tagged by BS0 and BS04 are the results of the present calculation. In the first one, we consider
a pseudo--vectorial coupling for the pion--nucleon interaction and in the second one
a mixed coupling (pseudo--vectorial plus pseudo--scalar) as described in the text.
The remaining curves result from calculations using the simplified deuteron
BS wave function of Ref.[11]\,(BSsm), the non--relativistic deuteron wave function of
Ref.[10] and the light--cone approach of Ref.[21]. \label{fig1}}
\end{figure}

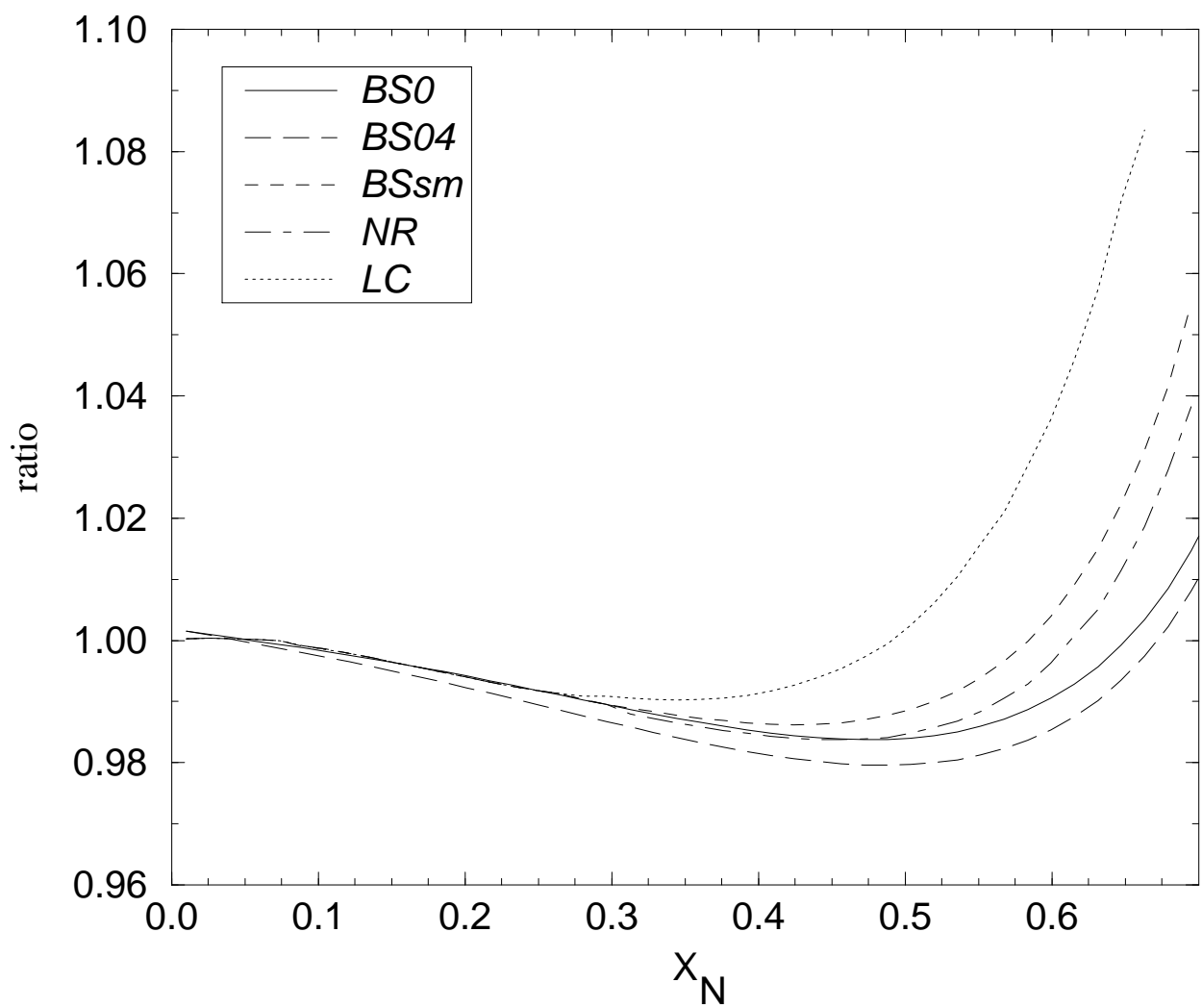
\end{document}